%% file: main.tex
\documentclass{ieeeaccess}
\usepackage{cite}
\usepackage{amsmath,amssymb,amsfonts}
\usepackage{algorithm}
\usepackage{algorithmicx}
\usepackage{algpseudocode}
\usepackage{graphicx}
\usepackage{textcomp}
\usepackage{float}
\usepackage{enumitem}
\usepackage[implicit=false,hidelinks]{hyperref}
\usepackage{textcomp}
\usepackage{multirow}
\usepackage{tikz}
\usepackage[final]{microtype}

\NewSpotColorSpace{PANTONE}
\AddSpotColor{PANTONE} {PANTONE3015C} {PANTONE\SpotSpace 3015\SpotSpace C} {1 0.3 0 0.2}
\SetPageColorSpace{PANTONE}%

\algnewcommand\algorithmicinput{\textbf{Input:}}
\algnewcommand\Input{\item[\algorithmicinput]}

\algnewcommand\algorithmicoutput{\textbf{Output:}}
\algnewcommand\Output{\item[\algorithmicoutput]}

\algnewcommand\algorithmicinitialize{\textbf{Initialize:}}
\algnewcommand\Initialize{\item[\algorithmicinitialize]}

\let\oldReturn\Return
\renewcommand{\Return}{\State\oldReturn}

\usepackage{circledsteps}
\pgfkeys{/csteps/inner color=white}
\pgfkeys{/csteps/outer color=none}
\pgfkeys{/csteps/fill color=black}

\def\BibTeX{{\rm B\kern-.05em{\sc i\kern-.025em b}\kern-.08em
    T\kern-.1667em\lower.7ex\hbox{E}\kern-.125emX}}
\begin{document}
\history{Date of publication xxxx 00, 0000, date of current version xxxx 00, 0000.}
\doi{EMPTY}

\title{SHIELD: An Adaptive and Lightweight Defense against the Remote Power Side-Channel Attacks on Multi-tenant FPGAs}
\author{
\uppercase{Mahya Morid Ahmadi}\authorrefmark{1},~\IEEEmembership{Graduate Student Member,~IEEE,},
\uppercase{Faiq Khalid}\authorrefmark{1}, ~\IEEEmembership{Member,~IEEE,},
\uppercase{Radha Vaidya}\authorrefmark{2},~\IEEEmembership{Student Member,~IEEE,},
\uppercase{Florian Kriebel}\authorrefmark{1},~\IEEEmembership{Member,~IEEE,},
\uppercase{Andreas Steininger}\authorrefmark{1},~\IEEEmembership{Member,~IEEE,},
\uppercase{Muhammad Shafique}\authorrefmark{3},~\IEEEmembership{Senior~Member,~IEEE}
}
\address[1]{Technische Universit\"at Wien (TU Wien), Vienna, Austria}
\address[2]{The University of California, Santa Barbara}
\address[3]{Division of Engineering, New York University Abu Dhabi (NYUAD), Abu Dhabi, United Arab Emirates}
\tfootnote{This work has been supported by the Doctoral College Resilient Embedded Systems which is run jointly by the TU Wien’s Faculty of Informatics and the FH-Technikum Wien.}


\corresp{Corresponding author: M. Morid Ahmadi (e-mail: mahya.ahmadi@tuwien.ac.at).}




\begin{abstract}
Dynamic partial reconfiguration has enabled multi-tenancy in cloud-based FPGAs, and such Infrastructure-as-a-Service (IaaS) has fueled remote computing in virtual accelerators. In a multi-tenant FPGA scenario, that allows sharing resources between different parties, certain key challenges need to be addressed to ensure the security of tenants, IPs, and data. Recently, it has been shown that malicious users can exploit FPGAs in remote side-channel attacks (SCAs) and extract confidential information from other tenants. Although logical separation ensures the integrity of designs, shared on-chip resources, e.g., power distribution network (PDN), have been exploited for remote attacks. Conventional SCA mitigation on the algorithmic level (masking and hiding) can address this threat in some cases. However, it requires significant effort to replace current library-based implementations and mappings. On the other hand, while bitstream checking techniques utilized by cloud providers can limit the insertion of attack sensors, they can not provide high accuracy in attack detection. Hence, an active on-chip defense mechanism is required to guarantee the confidentiality of execution for tenants.

Towards this, we propose a lightweight shielding technique utilizing ring oscillators (ROs) to protect applications against remote power SCA. Unlike existing RO-based approaches, in our methodology, an offline pre-processing stage is proposed to carefully configure power monitors and an obfuscating circuit concerning the resource constraints of the board. Detection of power fluctuations due to application execution enables the obfuscating circuit to flatten the power consumption trace. To evaluate the effectiveness of the proposed SHIELD, we implemented it on a Xilinx Zynq-7000 FPGA board executing an RSA encryption algorithm. Due to the SHIELD, the number of traces required to extract the encryption key is increased by 166x, making an attack extremely hard at run-time. Note that the proposed SHIELD does not require any modification in the target application. Our methodology also shows up to 54\% less power consumption and up to 26\% less area overhead than the state-of-the-art random noise-addition-based defense. 

\end{abstract}


\titlepgskip=-15pt

\maketitle

\input{Sections/Sec1_Introduction}
\input{Sections/Sec2_Background}
\input{Sections/Sec3_AttackModel}

\input{Sections/Sec4_Methodology}
\input{Sections/Sec5_MonitorDesign}
\input{Sections/Sec6_NoiseGenerator}

\input{Sections/ExperimentalSetup}

\input{Sections/Result}

\input{Sections/Conclusion}

%
\def\bibfont{\footnotesize} 
\bibliographystyle{IEEEtran}
\bibliography{main.bib}
\EOD

\end{document}

%% file: Sections/Sec1_Introduction.tex
\section{Introduction}
\label{sec:introduction}

 Due to high-performance and cost-effectiveness, the interest of adopting multi-tenancy in cloud-based computing platforms, like FPGA-based cloud servers, is increasing rapidly~\cite{jin_2020}\cite{Tutorial-20}. In an FPGA-as-a-Service (FaaS) computing model, resource sharing is inevitable, which raises several threats~\cite{faas}. For example, cloud-based FPGAs are deployed as virtual accelerators in cryptographic applications with secret keys, utilizing on-chip PUF-based keys against IP stealing attacks~\cite{SecureIP}. Furthermore, in machine learning applications such as deep learning, where techniques and algorithms are evolving swiftly, reconfigurable virtual accelerators are considered for performance enhancement~\cite{Tensor}. On the other hand, this idea needs to be cost-effective (e.g., in multi-tenant FPGAs) and secure to be offered by Cloud Service Providers (CSPs)~\cite{BNN}.  
 
 Recent studies show that these multi-tenant FPGAs are vulnerable to remote side-channel attacks (SCAs)~\cite{cpamap_2020} and covert channel attacks~\cite{Covert} due to their partial reconfiguration feature. For example, an attacker can exploit reconfigurability by inserting attack monitors and collecting the required set of traces for SCA. After a successful attack, the attacker reconfigures the FPGA to a benign application and removes attack monitors before being detected by the CSP or the target~\cite{Remote}. Based on this principle, researchers have developed several \textit{remote attacks} for multi-tenant FPGAs by exploiting the shared power distribution network (PDN)~\cite{zhao_2018}~\cite{schellenberg_inside_2018}, and the cross-talk effect between long wires~\cite{ramesh_fpga_2018}\cite{giechaskiel_leaky_2019}. Moreover, researchers also exploited these shared resources (e.g., PDN) to develop several \textit{fault-injecting attacks}, e.g., voltage-drop fault attacks using ring oscillators (ROs)~\cite{Gnad_fault-2017}~\cite{fpgahammer}, voltage-drop fault attacks without ROs~\cite{Sugawar-2019}~\cite{mahmoud_2019}~\cite{Hell2020}, and interrupting the power distribution network for aggressive power consumption~\cite{PoWaster_Provelengios}~\cite{Shutdown}. In short, these vulnerabilities limit the effective time window for detecting and mitigating SCAs for multi-tenant FPGAs. Hence \textit{it is imperative to develop effective and efficient run-time defense mechanisms for multi-tenant FPGAs.} 

\subsection{Limitations of the State-of-the-Art Defenses}
In a remote side-channel attack on FPGA cloud infrastructure, the users cannot have access to the hardware at the transistor level in order to modify the underlying platform's technology construction or deploy custom design hardware . Therefore, conventional ASIC mitigation techniques, such as dual-rail design~\cite{dual_rail} and supply current adjustments~\cite{Current}, are ineffective against this attack model. 
To mitigate and detect the above-mentioned remote SCAs, several defense mechanisms have been developed, e.g., hardware monitoring by the CSPs~\cite{gnad_checking_2018,FPGADefender,tian_fingerprinting_2020} to detect suspicious circuits in the bitstreams, the physical isolation of tenant designs~\cite{seifoori_closing_2020}\cite{luo2019hill}, run-time monitoring~\cite{runtime}, and power obfuscation using dummy hardware/software modules~\cite{Gnad_fault-2017}\cite{SCA}. A summary of state-of-the-art techniques is shown in Table~\ref{tab:WorksinSoA}. However, these defense mechanisms exhibit the following key limitations:

\begin{enumerate}[leftmargin=*]
 \item It is typically infeasible to restrict the usage of certain cells~\cite{gnad_checking_2018,FPGADefender,tian_fingerprinting_2020}, like ROs, because this can affect the performance of standard security primitive IPs, e.g., RO-based PUF and RO-based TRNG~\cite{tian2020cloud}. 
 \item Physical isolation~\cite{seifoori_closing_2020}\cite{luo2019hill} is effective in mitigating cross-talk-based and remote power SCAs\footnote{The bit value of transmitting on adjacent long wires can be exploited as a cross-talk side channel to obtain information from a victim.~\cite{giechaskiel_leaky_2019}}. However, due to design constraints and the inevitability of shared resources, physical isolation is not always feasible.
 \item Run-time monitoring~\cite{runtime} by the CSPs can detect suspicious behaviors, e.g., peaks in entering current, and can block malicious computations. However, these defense mechanisms are only applicable to active attacks (like fault-injecting attacks) and cannot detect passive attacks (like remote SCAs). 
 \item The dummy hardware/software modules~\cite{Gnad_fault-2017}\cite{SCA} can effectively obfuscate the power and delay behavior, which is a key feature in power-based remote SCAs. However, these modules come with high power and area overheads, computational cost, and complicated control unit. Moreover, the identification and control of vulnerable modules are challenging because they are tightly linked with the application-to-be-protected. 
\end{enumerate}

\begin{table*}[ht]
\label{tab:WorksinSoA}
\centering
\begin{tabular}{|c|c|c|c|c|c|c|c|}
\hline
\multirow{2}{*}{Technique} & \multicolumn{2}{c|}{Targeted side-channel attack} & \multicolumn{3}{c|}{Defense deployment point} & \multirow{2}{*}{\begin{tabular}[c]{@{}c@{}}Disclosure \\of design\end{tabular}} & \multirow{2}{*}{\begin{tabular}[c]{@{}c@{}}Type of defense\end{tabular}} \\ \cline{2-6}
   & Cross-talk & Power  & Cloud & Vendor & User      &      &   \\ \hline
\cite{FPGADefender},~\cite{krautter_mitigating_2019} &  X     & X       & X    &  -   &   -  &  -  & Passive   \\ \hline
\cite{seifoori_closing_2020} ,~\cite{luo2019hill} & X   & -  &  -   & X    &  -   &  X  & Passive  \\ \hline
\cite{Constant_Power_2011}   & -  & X    &  -   & -  & X    &  -   & Passive      \\ \hline
\cite{krautter_active_2019}  & -   & X    & X     & - & X    & -    & Active       \\ \hline
SHIELD    & -  & X    & X     &  - & X    &   -  & Active    \\ \hline
\end{tabular}

\caption{\centering Comparison of mitigation techniques against side-channel attacks to remote FPGAs}
\label{Tab:Comparison}
\end{table*}

\subsection{Motivational Case Study and Targeted Research Challenges}
\begin{figure}
 \centering
 \includegraphics[width=1.0\linewidth]{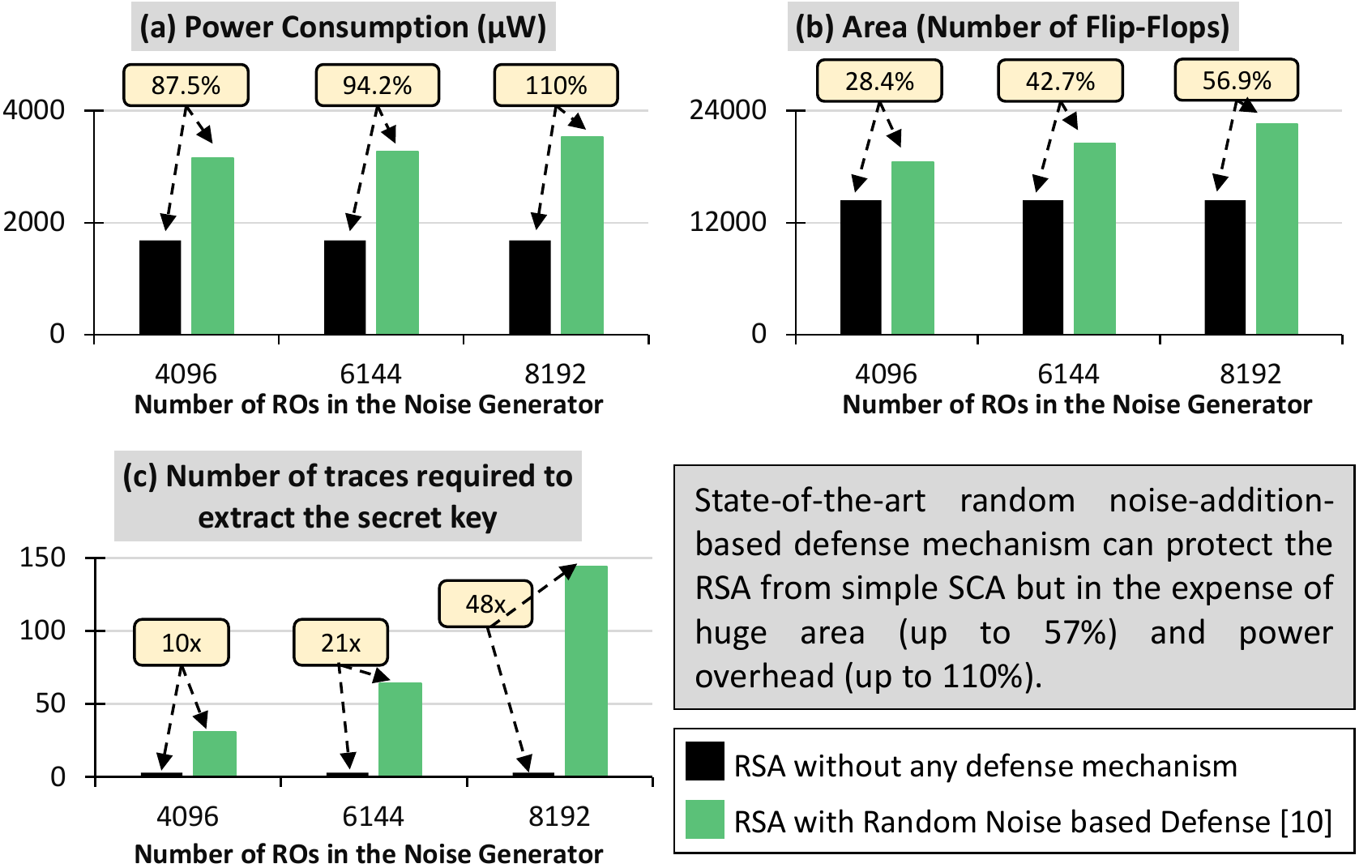}
\caption{RO with random noise-addition-based defense mechanism~\cite{zhao_2018} to protect RSA from power-based SCAs. Note, in this analysis power consumption overhead is calculated using Vivado Power Estimator and the area overhead is reported by Vivado Place~\&~Route tool for different sizes of noise generators (number of ROs) with respect to the number of utilized Flip Flops. }
\label{fig:MA} 
\end{figure}

To address the above-discussed limitations, recently, researchers have developed several mitigation techniques that actively inject noise into power traces for obfuscating power behavior and use random noise-generating modules (e.g., ROs) between trusted and untrusted designs. These techniques typically use always-on ROs to insert the random noise, but they exhibit large area and power overheads. To highlight such techniques' limitations, we implemented the state-of-the-art random noise addition-based defense~\cite{zhao_2018} to protect RSA running on the Xilinx Zynq-7000 FPGA board from simple power-based SCA. As shown in Fig.~\ref{fig:MA}, our analysis demonstrates that random noise addition-based defense  increases the number of traces required to extract the key of RSA encryption by 48x while exhibiting 56.9\% area overhead and 110\% power overhead. 

To address these limitations, recently, researchers have proposed a technique~\cite{krautter_active_2019} that introduces controlled noise instead of random noise. The key challenge in such defenses is to design an appropriate power monitor that can sense the power fluctuations, and activate the noise accordingly. Recently, this challenge is addressed by using Time-to-Digital Converters (TDC) as the power sensors~\cite{krautter_active_2019}. Although TDCs are sensitive to power fluctuations and provide high-resolution power values and high accuracy, they consist of a complex and long chain of LUTs, registers, and multiplexers. Hence, they exhibit large area overhead (e.g., it led to 100\% area overhead for a defense technique applied to an AES implementation, as reported in~\cite{krautter_active_2019}) and huge power overhead (e.g., 50\% overhead with respect to AES power, as reported in~\cite{krautter_active_2019}). \textit{Therefore, there is a dire need for a generic methodology to design an appropriate power-obfuscation mechanism for on-demand defense activation.}   

\textbf{Open Research Challenges:} The above-mentioned limitations of noise-addition defenses lead to the following key research challenges:

\begin{enumerate}[leftmargin=*]
  \item How to find an appropriate configuration of the power monitor to add a controlled noise for a given application on an FPGA platform?
  \item How to reduce the complexity of the sensors in the power monitor while reducing the respective area and power overheads?
\end{enumerate}

\subsection{Our Novel Concept and Contributions}
\begin{figure}
 \centering
 \includegraphics[width=\linewidth]{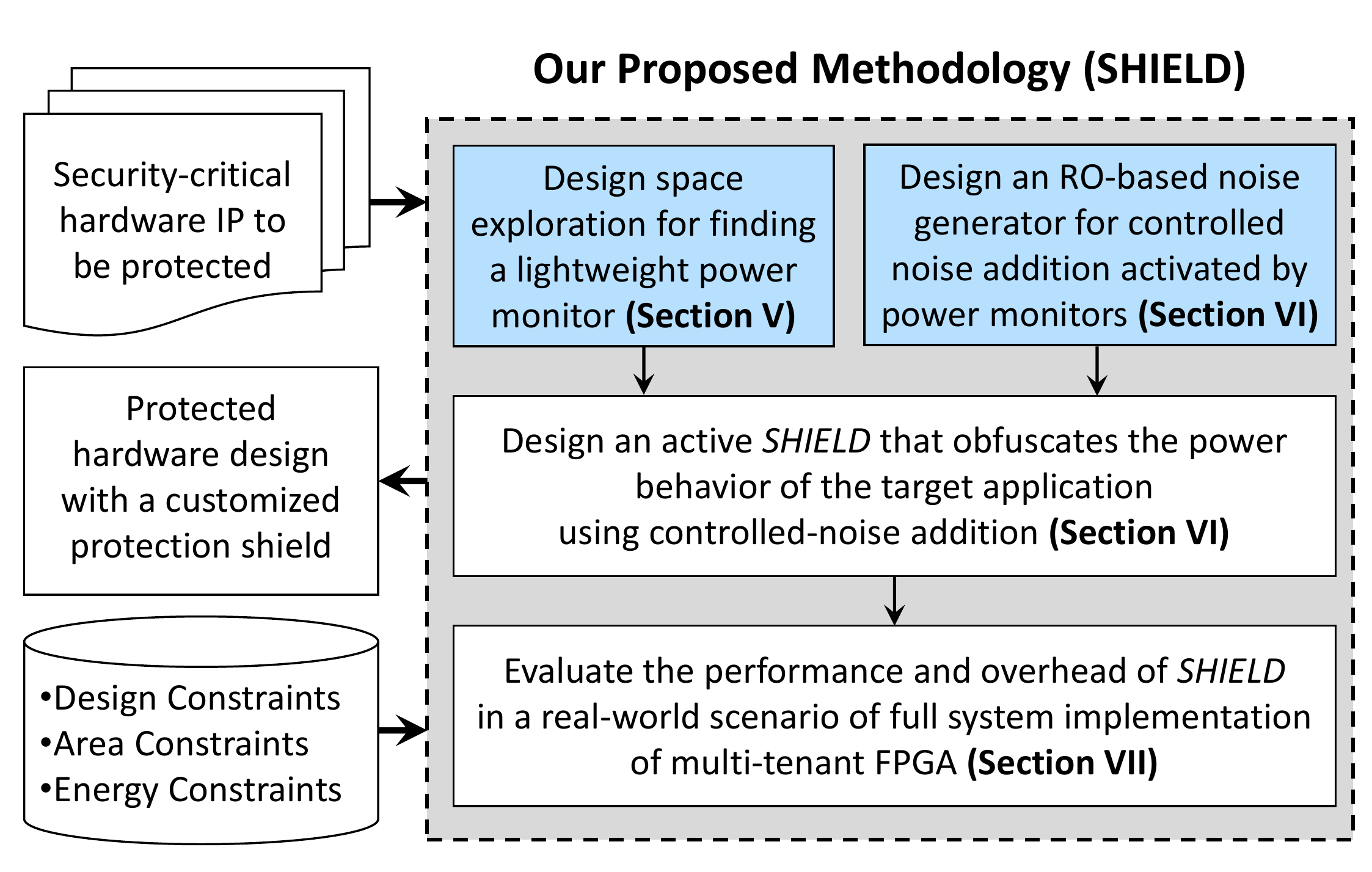}
 \caption{Overview of our novel contributions (highlighted in blue.)}
 \label{fig:Contributions} 
\end{figure}

To address the above-mentioned challenges, we propose a novel design methodology, SHIELD, that utilizes RO-based counters as sensors in power monitors of the controlled noise addition-based defense (see the overview of the methodology in Fig. ~\ref{fig:Contributions}). In summary, the key novel contributions of this paper are the following:

\begin{enumerate}[leftmargin=*]
  \item \textbf{Lightweight Power Monitor (Section~\ref{sec:Power-Monitors}):} Unlike the state-of-the-art random noise addition-based defense, we propose a novel methodology that performs a design space exploration to design an appropriate lightweight power monitor using RO-based counters as power sensors to enable controlled noise addition. The reason for choosing the RO-based counters is that they provide better spatial coverage and are less complex (consisting of basic logic gates), making them suitable for lightweight power sensors.
    \item \textbf{Optimization of Noise Generators (Section~\ref{sec:Noise_Generator}):} To obtain the appropriate size and configuration of the RO-based noise generators, we propose a methodology that considers the victim's application, target board, and its power behavior to optimize the size of ROs and the appropriate number of ROs in the noise generator for better accuracy and less area and power overhead.
\end{enumerate}

To evaluate the effectiveness of our proposed defense, we designed an application-specific power obfuscation module against remote power attacks. We implemented this setup on the Xilinx Zynq-7000 FPGA board and evaluated it by power-based SCA~\cite{ors2003power}. The experimental results show that this defense module, designed using our methodology, effectively obfuscates the power behavior of RSA while exhibiting $\approx{58\%}$ power overhead and $\approx{30\%}$ area overhead. Note that these power and area overhead are $\approx{50\%}$ less as compared to the state-of-the-art random noise-addition-based defense~\cite{zhao_2018} against power SCA. 

%% file: Sections/Sec2_Background.tex
\section{Background}\label{Background}
In this section, to better understand the proposed methodology, we briefly discuss the basic principles of RO-based power monitoring and the implementation of  a simple power SCA on the Rivest–Shamir–Adleman (RSA) encryption algorithm which is one of the most-used public-key cryptosystems.

\subsection{Basic Principles of RO-based Power Monitoring}
The power consumption of CMOS circuits consists of dynamic power ($P_{dyn}$) and static/leakage power ($P_{leakage}$). The frequency of the input and switching activity of the different nodes due to computations in the running application can affect the dynamic power, and it is modeled as: 
\begin{equation}
    P_{dyn}= \alpha \times V_{DD}^{2} \times f_{SW} \times C
    \label{eq:p_dyn}
\end{equation}
The fluctuations in dynamic power consumption are directly related to the switching activity of the circuits; thus, these fluctuations can be used to estimate the input data and respective computations. The monitoring of dynamic power consumption requires direct access to the power rails of the circuit, but it is impracticable to have direct access in real-world scenarios. However, shared power rails between different hardware modules can be monitored to sense these power fluctuations. In an SoC, the PDN manages the power supply to the different blocks/components. An ideal PDN provides steady voltage and current and is resilient to fluctuations. However, the voltage regulators in a PDN cannot provide steady voltage because they are sensitive to switching activities in hardware modules ($I(t)\propto f_{sw}$). Mathematically, a PDN can be modeled as an RLC circuit using the following set of equations.

\begin{equation}
    \Delta V_{drop,R}=I(t) \times R
    \label{eq:v_drop}
\end{equation}
\begin{equation}
    \Delta V_{drop,L}=L \times \frac{dI(t))}{dt}
    \label{eq:v_drop2}
\end{equation}

Different computations and variations in operating conditions are reflected in the dynamic power consumption that results in a change in the supply current and voltage ($P= V I$).  Since the PDN is not ideal, the change in $I$ and $V$ can be measured depending on the power. Thus, these changes in $I$ and $V$ can be used to estimate the computations based on the input data.  

By exploiting these non-idealities in the PDN, researchers have developed RO-based power monitors in FPGAs because the frequency of the RO is sensitive to voltage variations in the PDN. An RO consists of an odd number of inverters, and the output of the last inverter is fed back to the first inverter. To enable the array of ROs, typically, an AND gate (where one input is the enable signal and another one is a feedback signal) is used at the input of the RO. The oscillation frequency of the RO is tightly coupled with the voltage variations~\cite{barbareschi2017}). By adjusting the number of inverters, the resolution of detecting these voltage variations in the PDN can be changed, and it can be modeled as:

\begin{equation}
    f_{RO}~\propto k* V~(x,~y) + f_{0}
    \label{eq:freq}
\end{equation}

Where the $f_{0}$ and $k$ are constants and $V(x,y)$ is the voltage at the RO's location in the FPGA.

\begin{figure}[!t]
    \centering
    \includegraphics[width=1\linewidth]{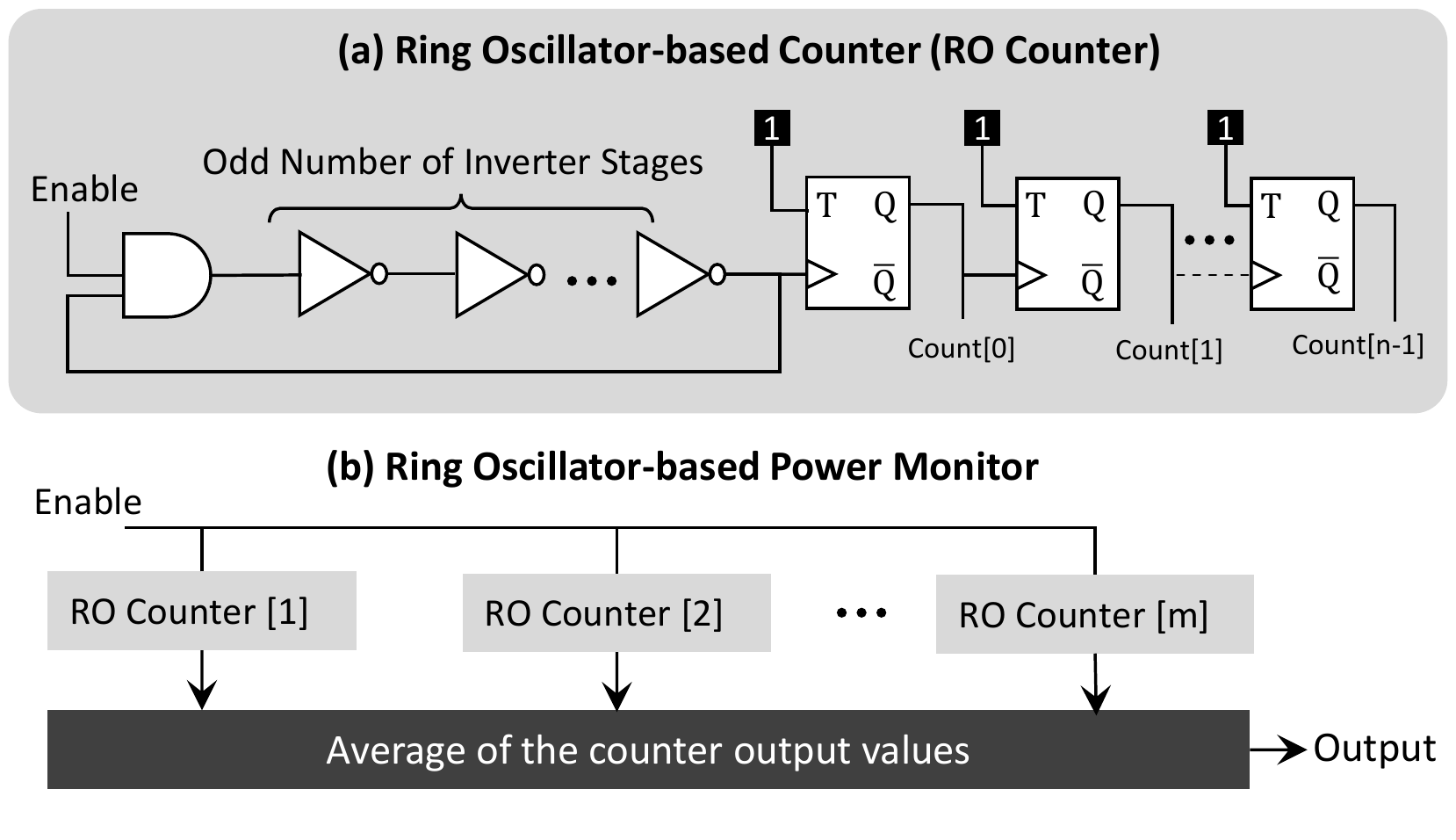}
  \caption{Basic design of RO counter and RO-based power monitor.}
    \label{fig:RO} 
\end{figure}

To measure the RO frequency, typically, a T-Flip-Flop (T-FF) is used as a counter that counts every oscillation in the RO, shown in Fig.~\ref{fig:RO}(a). As shown by Giechaskiel et al. in~\cite{loop}, in the RO-based power monitors, using the T-FF evades the combinational loop detector of FPGA compilers and prevents the elimination of circuits by cloud providers while limiting the frequency of the RO. For the proof-of-concept, we adopt the standard design of ROs, which can be updated to non-combinational loop ROs in future works. Since the oscillation frequency is greater than the Output register's capturing frequency, a chain of T-FFs is required to count the oscillations, as shown in Fig.~\ref{fig:RO}(a). To improve the precision, multiple RO-based counters are used to estimate the fluctuations in different locations of the SoC, as shown in Fig.~\ref{fig:RO}(b). Then, the final value is computed by taking the average of all RO's outputs.

\subsection{Security Vulnerabilities in RSA w.r.t. Remote Power SCA}
RSA is one of the most-used public-key encryption algorithms, and its effectiveness is determined by the size of the key and the time-to-break using brute force. The typical size of the public key ranges from 1024 to 4096 bits. Since it is an effective and powerful encryption scheme, the computational demand is high. For example, the computations required for determining the modular exponentiation of larger numbers are very costly to be implemented in software (in terms of execution time and power). Therefore, modern secure systems are updated to hardware accelerators. In RSA, one of the common techniques to compute modular exponentiation is using the algorithm implemented as Square \& Multiplication Module (SMM), as shown in Algorithm~\ref{algo:RSA_ME}. This algorithm has a  distinct computational pattern with respect to the bit-wise values of the public key. For example, in the conditional loop of Algorithm~\ref{algo:RSA_ME}, a multiplier is enabled if the value of the public key is `1'; hence it consumes more power. However, when the value is `0', the multiplier is not used; hence consumes relatively less power. These properties make the RSA vulnerable to remote power SCA.

Moreover, other techniques are proposed in the literature to enhance the resiliency of RSA against SCA, i.e., constant-time RSA~\cite{bearssl}. These secure implementations of RSA are proposed to mitigate timing side-channel attacks~\cite{cachebleed} and it has been shown that constant-time designs can not protect applications effectively against power side-channel attack~\cite{SecureRSA}, especially in a hardware implementation; therefore, it does not address power side-channel attacks SHIELD protection is orthogonal to constant-time implementation.
Therefore, in this work, our primary goal is to develop a lightweight solution that can obfuscate the impact of distinct computations on the power behavior of critical applications. As a use case, we choose to protect an RSA algorithm.

    \begin{algorithm}[!t]
    \footnotesize
    	\caption{\footnotesize{\textbf{:} Modular Exponentiation using Square-and-Multiply in RSA}}
    	\label{algo:RSA_ME}
    	\begin{algorithmic}[1]
    		\Input $M$: Plaintext,~~~$d$: Private Key Exponent,~~~$N$: Modulus of RSA
    		\Output $R$: Modular exponentiation of M
    		\Initialize $R \gets 1$, ~~~$S\gets M$
    		\For{$i=0~to~n-1$}
            \If {$d~mod~2~==~1$} 
            \State $R~=~R*S~mod~N$
            \EndIf
            \State $S~=~S*S~mod~N$
            \State $d~=~d >> 1$
            \EndFor
            \Return R
    	\end{algorithmic}
    \end{algorithm}

%% file: Sections/Sec3_AttackModel.tex
\section{Adversary Model}\label{Model}
We assume the most widely used threat models for multi-tenant FPGA-based cloud services~\cite{krautter_active_2019}, namely that the attacker can be any prospective tenant and exploit the shared resources of a multi-tenant FPGA, even in the situation of separate partial bitstreams. While the attacker and victim share the same FPGA, they are logically and physically separated in this scenario, which is assured by the use of reputable FPGA vendor design tools. Additionally, we presume the attacker is familiar with the target application that is executing on the victim's partial area.
This assumption is based on the categorization of the victim's application in remote side-channel circumstances by works such as~\cite{Classify}.
In this article, we suppose that the FPGA has simply the attacker and victim.
While this assumption may be made in practice, given the nature of noise detection, the worst-case situation is for the victim to obfuscate his power trace. In multi-user circumstances, the victim tenant can continue to benefit from our proposed defense, SHIELD, as an add-on to the target application's security, without the need for accelerator designers to be aware of side-channel leakages.
Additionally, cloud service providers may use SHIELD to provide a lightweight security supplement for mission-critical applications.

%% file: Sections/Sec4_Methodology.tex
\section{Our Novel Methodology to Design an Adaptive and Lightweight SHIELD against Remote SCA}
\label{Methodology}
\begin{figure}
    \centering
    \includegraphics[width=\linewidth]{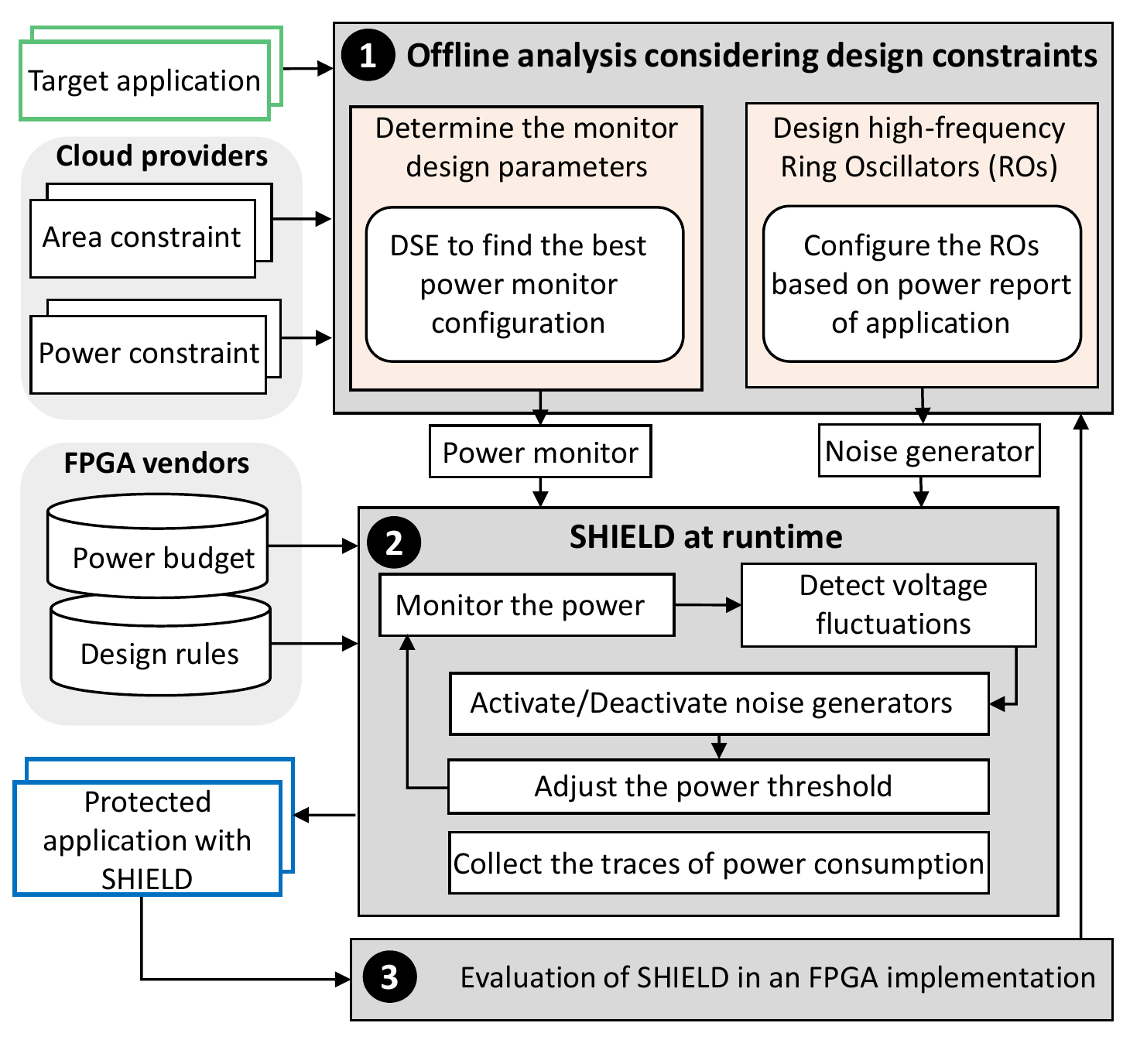}
    \caption{An overview of the proposed methodology to design an adaptive and lightweight SHIELD against remote SCA on multi-tenant FPGAs. The novel steps are highlighted with blue boxes and the yellow boxes are the input and output of the methodology.}
    \label{fig:Methodology}
    \end{figure}

\begin{figure}
    \centering
    \includegraphics[width=\linewidth]{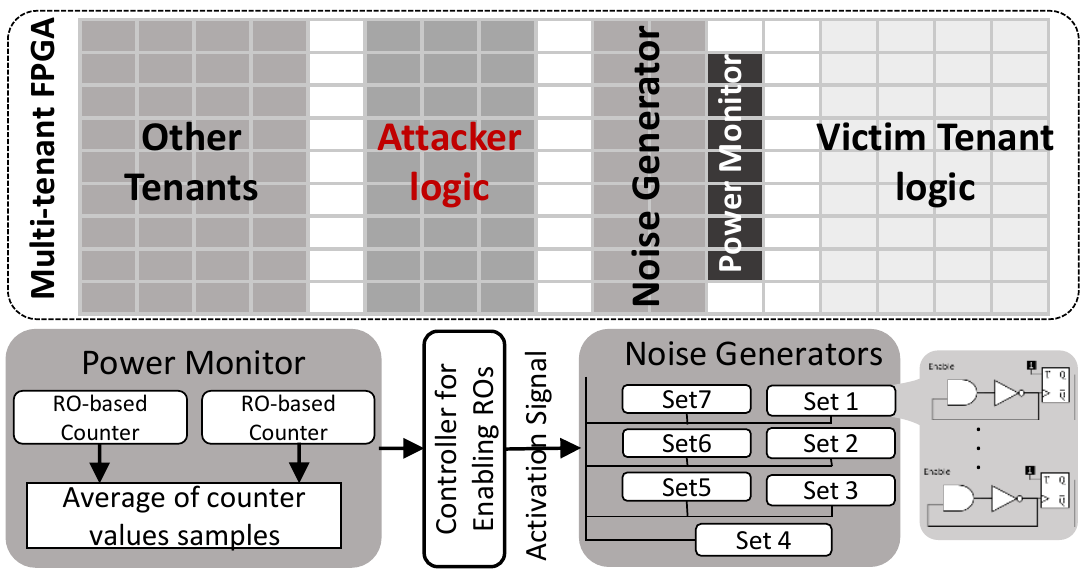}
    \caption{The scheme of SHIELD utilizes RO-counters for power monitoring and noise generator. To design the SHIELD, we have adopted the state-of-the-art design technique for high-speed RO-counters, as shown in Fig.~\ref{fig:RO}.}
    \label{fig:scheme} 
\end{figure}

Fig.~\ref{fig:Methodology} shows a detailed overview of our proposed methodology. SHIELD employs the following three steps to design a lightweight defense against remote SCA:
\begin{enumerate}
    \item In the first step, we explore different design options for configuring the power monitor and analyze the options based on design constraints, area constraints, and energy constraints (see step 1 in Fig.~\ref{fig:Methodology}). This design space exploration (DSE) aims to choose effective and lightweight designs for power monitors. The design configuration set used in the DSE consists of the size in terms of the number of RO counters, the placement of the power monitor, and the sampling frequency of the power monitor. The implementation details of the power monitor and noise generators are presented in the subsequent sections. 
    \item In the next step, the power monitor and an appropriate number of noise generators are used to design an adaptive SHIELD while considering the power budget and design rules from the FPGA-vendor (shown in Fig.~\ref{fig:scheme}). In SHIELD, the power monitor observes the power fluctuations in the PDN, and after detecting and analyzing the voltage fluctuations, it enables the noise generators with a specific number of power monitors based on the states of detection. These RO-based noise generators flatten the power consumption traces to obfuscate the computational patterns.
   \item The design of SHIELD is customized per application for the maximum power consumption of the target hardware to avoid unnecessary overhead. Moreover, in our methodology, it is evaluated for its protection capability and area/power overheads in a power analysis attack. 
\end{enumerate}

%% file: Sections/Sec5_MonitorDesign.tex
\section{Power Monitor Design}\label{sec:Power-Monitors}
In this section, we explain our novel methodology for designing power monitors in SHIELD. The voltage fluctuation inside the fabric can be estimated by designing propagation delay sensors such as Ring Oscillator (RO) sensors~\cite{high-speed} or Time-to-Digital Converters (TDC) based sensors~\cite{TDC}. In the design of SHIELD, we adopt RO-based monitors. First, we explain how physical parameters are selected and configured. Next, we show the results of a design space exploration, and after that, we show how we extract the key in this process.
\subsection{Ring Oscillator-based Power Monitor}
In the first step of SHIELD design, we propose an offline application-specific methodology to select the best configuration of the power monitor for the targeted applications. As shown in Section~\ref{Background}, ROs are proposed in the literature to be used as power sensors. In this work, we chain a series of RO counters to build up the power monitor. As shown in Fig.~\ref{fig:RO}, the average of {\it m} stages of ROs is collected as a sample in power trace to compensate for the effect of errors in counters. To simplify the {\it average} w.r.t., 
the function of division, {\it m} is always chosen as a power of {\it 2}. \par

The first parameter in the design of the power monitor is the size of the monitor. For each RO counter, we are employing {\it n} Flip-Flops with a fixed number of inverter stages, each triggers a counter. The studies on the design of power monitors show that the sensitivity of the monitor depends on its spatial proximity of it to the target application. To analyze the location parameter, we experimented with three locations to place two RO-based power monitors in the FPGA. Using this experiment we analyze the relation of monitor location by comparing the difference of the outcome of two monitors.

\begin{figure}
 \centering \includegraphics[width= 1.0\linewidth]{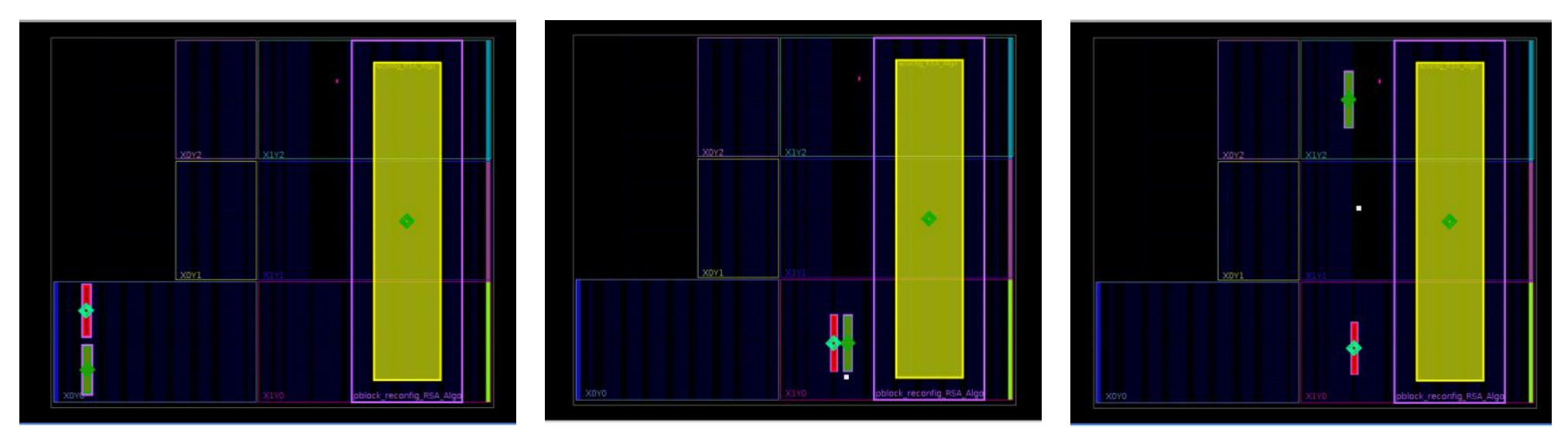}
 \caption{Placement Scenarios for power monitors. In left-most ROs in power monitors are placed spatially far, in the center the ROs are next to each other, and in the right-most, they are at close distance to RSA but far from each other.}
 \label{fig:placement}
\end{figure}

In Fig.~\ref{fig:placement}, this setup has been shown. The analysis on the bit-errors for 200 traces in each placement shows that the resolution of monitors increases when they are implemented in the closest distance to the leaky components of the target application, in this case, RSA modular multiplication modules.\par
Next, we analyze the number of ROs in each power monitor. For each monitor, we are employing a chain of {\it n} FFs, as a counter with the frequency of $f_{ref}$. The number of inverters limits the frequency of RO and the number FFs in an RO-counter defines the resolution of RO in capturing the power fluctuations. The trade-off between RO frequency in sampling the power traces and the accuracy of the power monitor is such that for higher accuracy, a lower frequency is required to have higher resolution sampling of cycle counts in RO. This frequency depends on the power fluctuations the monitor needs to observe, which is directly related to power consumption in the target application. The sampling frequency must be high enough not to miss any targeted voltage-drop. Zhao et al. in~\cite{zhao_2018} mention that the frequency of ROs follows the frequency of the system with $f_{RO}=C_{RO}*\frac{f_{Ref}}{C_{ref}}$, where the RO runs until the reference counter reaches a pre-determined sampling cycle count ${C_{ref}}$, then the ${C_{RO}}$ is read from registers (considering the quantization error in phase shift of two consecutive clocks). Following this, we analyze the accuracy of our power monitors with two frequencies, 10 MHz, and 100 MHz.\par
\begin{figure}
 \centering
 \includegraphics[width= 1.0\linewidth]{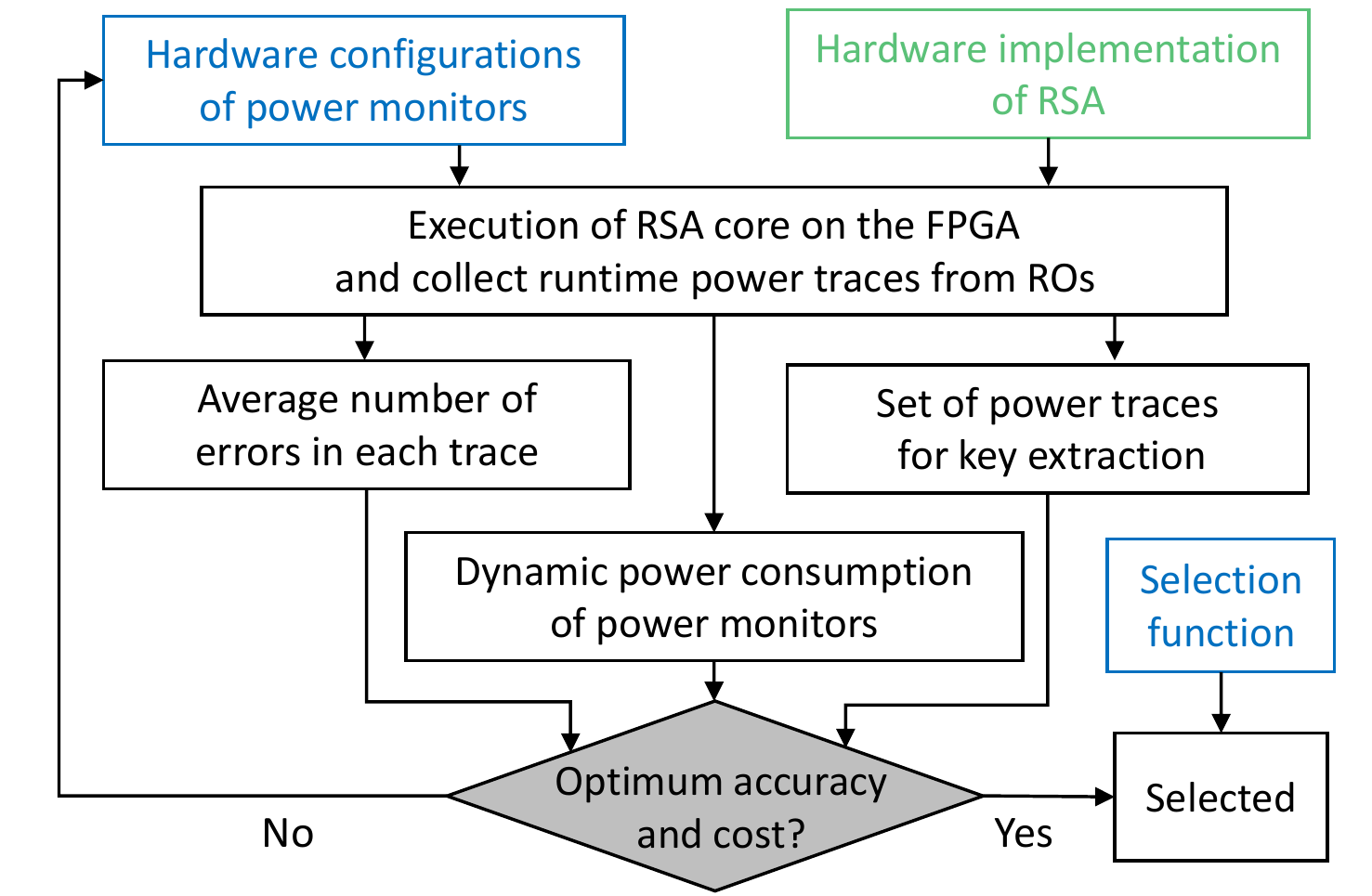}
 \caption{Design Space Exploration of the physical configuration for the Power Monitor.}
 \label{fig:Cost}
\end{figure}

 
The process of choosing a configuration fixes each parameter at a time and calculates a cost function considering the normalized values of overhead and accuracy of the monitor iteratively. In Fig.~\ref{fig:Cost}, the process of the power monitor configuration is shown. The output of this analysis provides an application adapted power monitor, which has balanced the trade-off between accuracy and overhead.\par

\subsection{Design Explore Results}
By evaluating physical parameters in the configuration of power monitors, we have explored the design space. The goal of DSE is to find the appropriate design choices for power monitors with respect to the average number of incorrectly detected key bits and the number of traces required to extract the key. 

\begin{figure}
\begin{center}
\includegraphics[width= 1\linewidth]{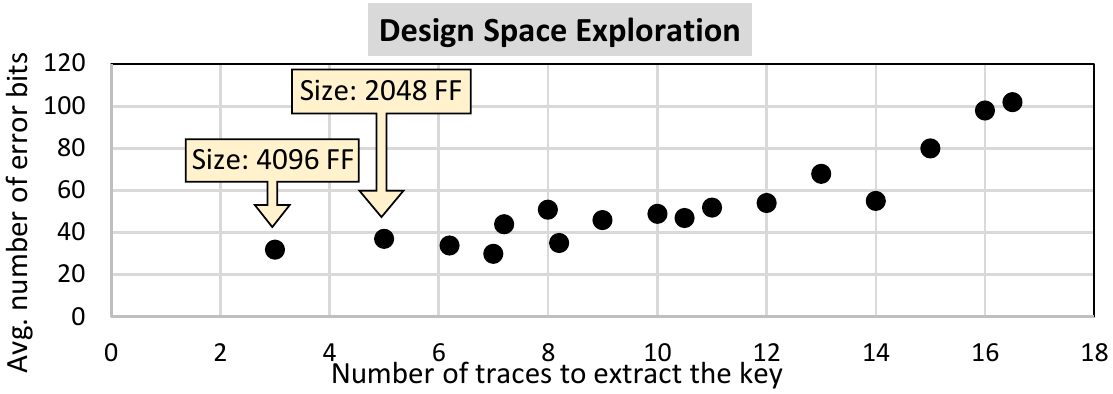}
\end{center}
\caption{Results of DSE for Power Monitor design.}
\label{fig:DSE} 
\end{figure} 
In Fig.~\ref{fig:DSE} we show different scenarios for three placement strategies (far, close1, close2), two frequency choices (10 MHz, 100 MHz), and, three numbers of RO counters (16, 32, 64). Although it has been shown by literature that increasing the number of RO counters improves the accuracy of detection, the power constraints of the chip limit the possible number of employed ROs. Since the 2 best design choices (close2, 10, 64) and (close2, 10, 32) are showing similar accuracy results, we choose the one with relatively 50\% less area overhead for the number of used FFs.

\subsection{Key extraction}
 In this section, we explain the principles of the key extraction algorithm. 
The principle of a side-channel attack is based on data-dependent execution flow. As mentioned in Section II, in our target application, RSA, the SM module is leaking the bits of the key. In the conventional implementation, SM decomposes the exponent $d$ into the sum of successive powers of 2 (i.e., in binary). Therefore, equation (\ref{eq:RSA}) can be computed using the following equation.

 \begin{equation}
 M=\prod_{i=0}^{n-1}(C^{2^i})^{d_{i}}~mod~N
 \label{eq:RSA}
 \end{equation}
 where $n$ is the length of decryption exponent $d$ and $d_i$ is the bit $i$ in $d$. (It has been mentioned that M is the message, d is the private key and C is the ciphertext).
 \par
 
\begin{figure}
 \centering
 \includegraphics[width= 1.0\linewidth]{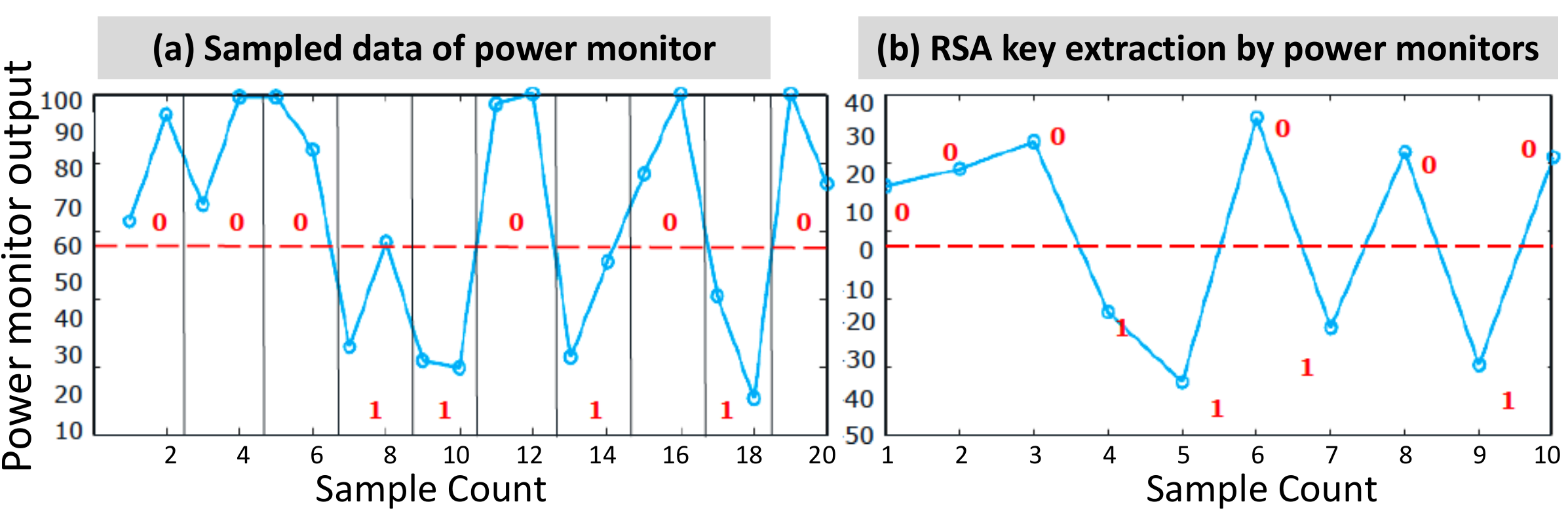}
 \caption{A sample of Power Monitor output for 10 bits of the key.}
 \label{fig:Key}
 \end{figure}

  As shown in Fig.~\ref{fig:Key}, then by calculating the average of the sampled data for each time slot in a trace, and comparing it to an adaptive threshold, we can detect peaks and valleys in the trace. The increased power consumption in the victim's application due to processing key-bit '1', leads to a drop in the share of voltage for the power monitor, and a lower number of oscillations is recorded. On the other hand, processing key-bit '0' results in a higher share of voltage in the power monitor and causes a larger than a threshold number of oscillations.
 
%% file: Sections/Sec6_NoiseGenerator.tex
\section{Noise Generator}\label{sec:Noise_Generator}

To control the noise generators, a drop-detection and a noise-activation process are required to control the level of noise and activate the noise generators accordingly. To detect the peaks and valleys in the trace of the monitor output, an initial threshold is set. The analysis shows that by activating each set of noise generators, the mean of dynamic power increases. To detect the next peak correctly, the controller increases the threshold and consequently will reset it to the initial value when all noise generators are deactivated. 
\begin{figure}[ht]
    \centering
    \includegraphics[width=\linewidth]{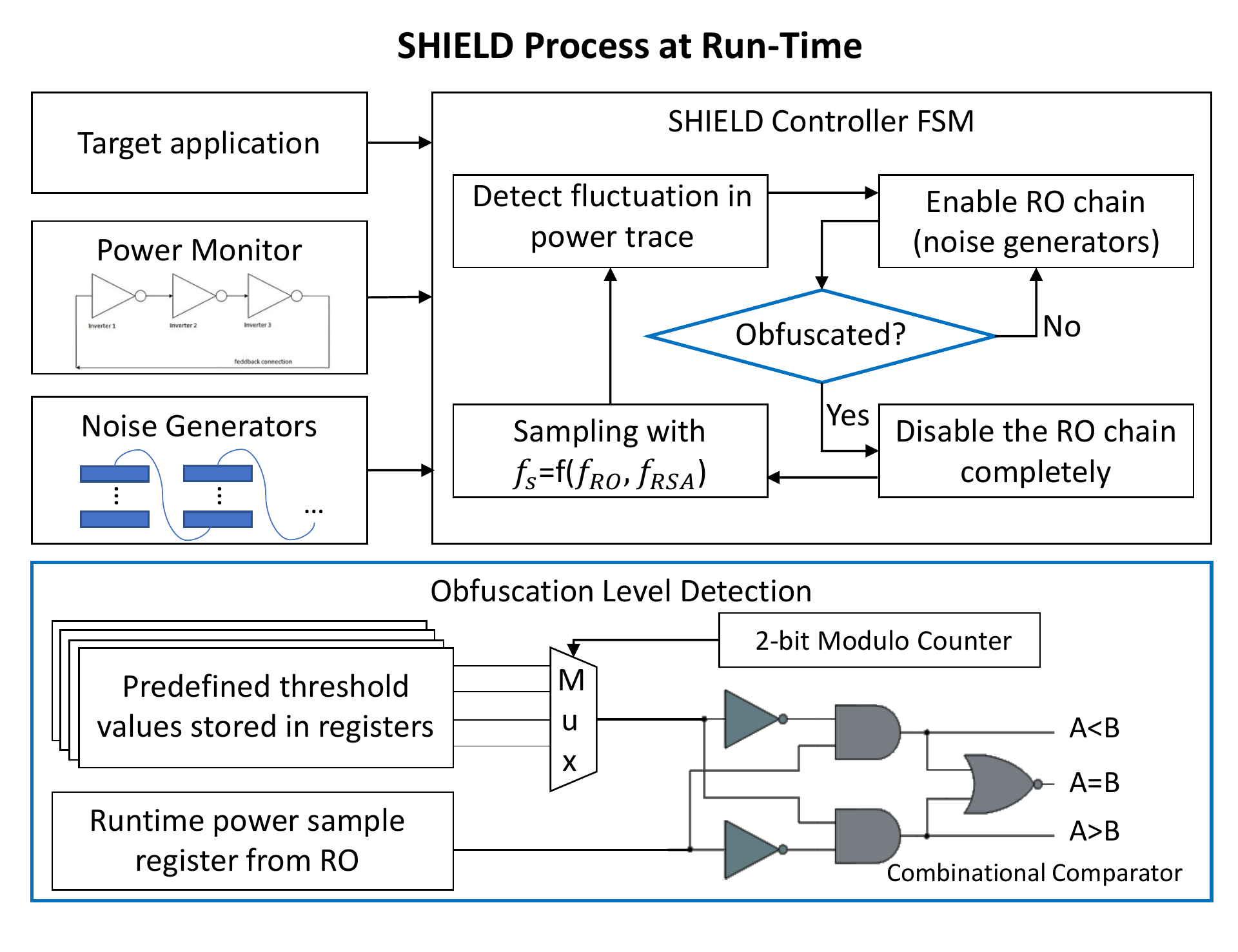} 
    \caption{Methodology of SHIELD operations at run-time.}
    \label{fig:OnlineShield}
\end{figure}

After detecting a peak in the power monitor's oscillation frequency (bigger share of power used by RSA to process key bit 1), a strategy is required to activate the noise generators. In Fig.~\ref{fig:OnlineShield}, the methodology of SHIELD in run-time is shown. Following the voltage drop detection, the activation process starts and activates the first set of ROs to increase the observed $P_{dyn}$, which increases the threshold level for measured output of power monitor to distinguish voltage-drop accordingly. By receiving the next measurement above the threshold in the output of ROs in monitors, the controller increases the number of active RO sets, which results in a higher level of noise introduced to the power trace. The maximum power consumption of the RO set in this module is proportionally half of a modular multiplier module in an RSA algorithm, following a design guideline of the maximum power budget of the SoC provided by the FPGA vendor. Therefore if the multiplication module is enabled while the noise generator is running, it causes the voltage-drop in all monitors, including the monitor ROs. By observing the first occurrence of values less than the threshold or the sample after activating all ROs in the output of power monitor, the controller deactivates the whole noise generator, as shown in Fig.~\ref{fig:OnlineShield}. This activation strategy obfuscates the power consumption related to the bits of the key. Furthermore, the increase in $P_{dyn}$ decreases the share of the total power of the RSA application. 
It can increase the attacker's effort to quantify the power trace to extract the key and add more error bits to the extracted key. The Noise generator is designed to consume maximally the amount of power used by the multiplier component when a key-bit '1' is processing.

%% file: Sections/ExperimentalSetup.tex
\section{Experimental Results} \label{Results}
To illustrate the effectiveness of the proposed controlled noise addition-based defense, we  evaluated it for RSA running on a ZedBoard and compare it with the random-noise addition defense mechanisms~\cite{zhao_2018}.  
\begin{figure}[ht]
    \centering
    \includegraphics[width=1\linewidth]{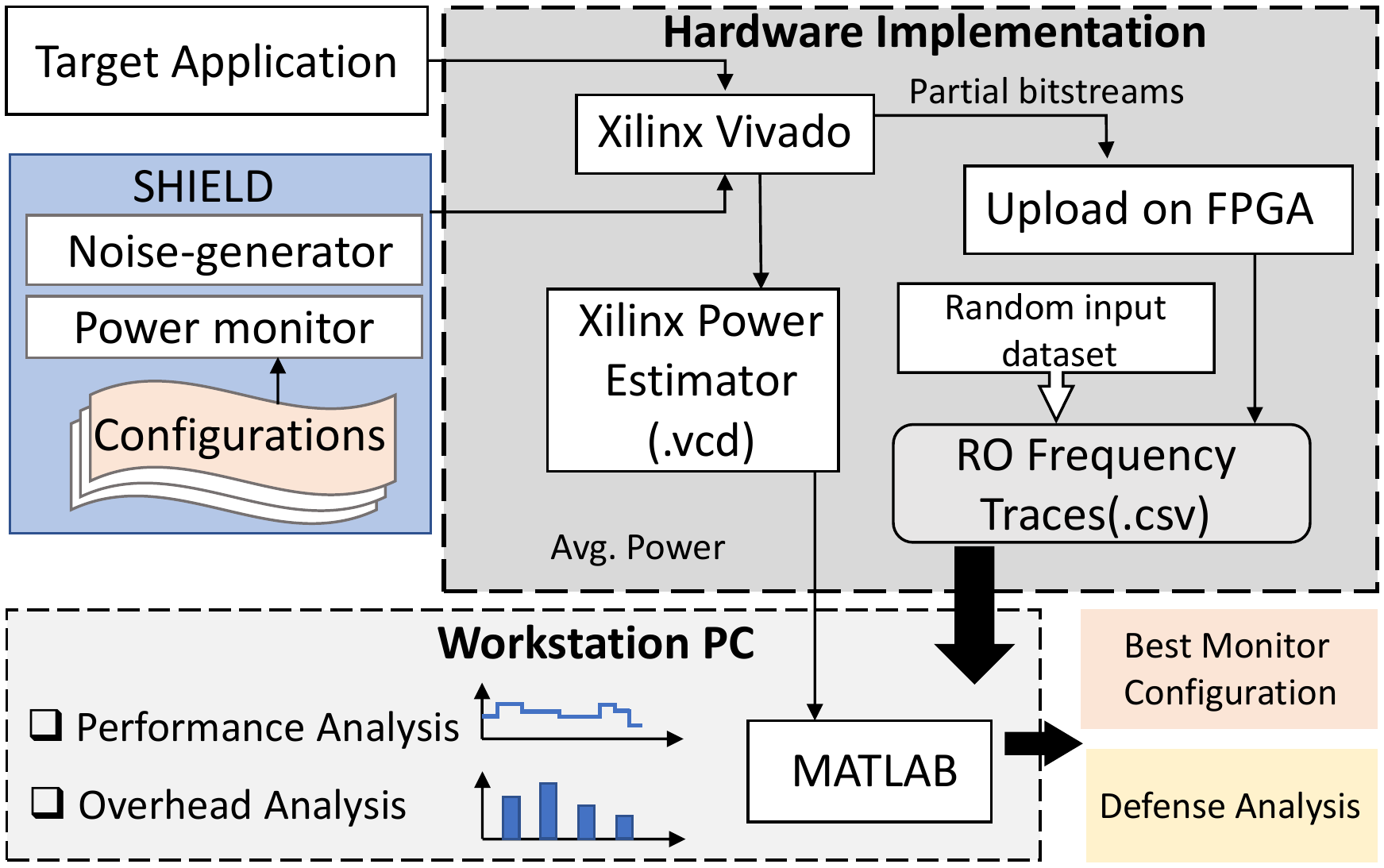}
    \caption{Tool-Flow diagram for SHIELD methodology.}
    \label{fig:Toolflow}
\end{figure}

 \begin{figure*}[hpt]
      \centering
      \includegraphics[width=\textwidth]{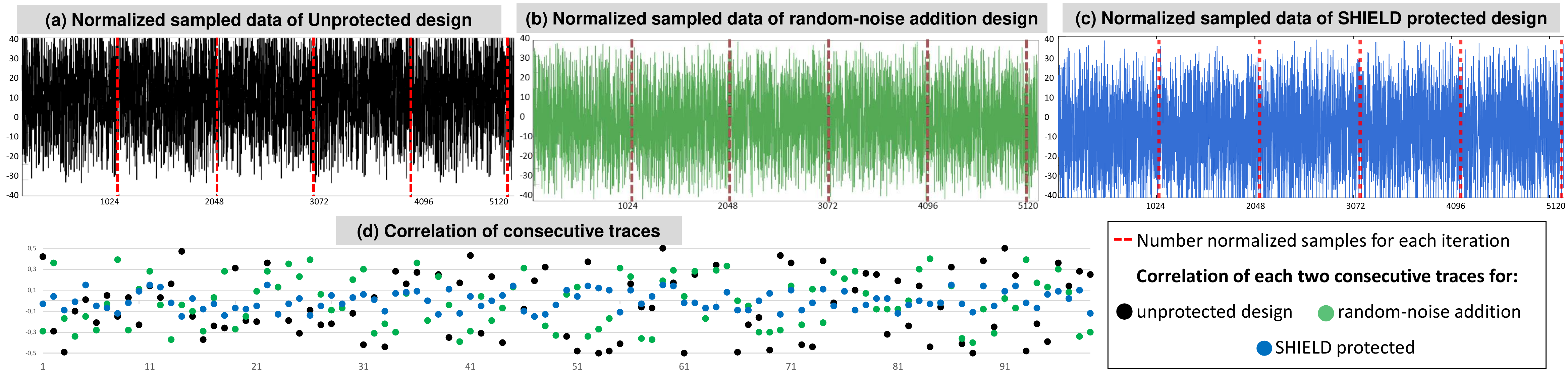} 
      \caption{Correlation analysis of SHIELD methodology in comparison of unprotected design and random noise addition-based defense. In (a), (b) and (c), the normalized sampled data for five consecutive traces of RSA running on an unprotected design, for random-noise addition and for SHIELD are shown, respectively. (d) The correlation of coefficients for each pair of consecutive traces (corr(trace n, trace n+1)).}
      \label{fig:Corr}
  \end{figure*}
 
\subsection{Experimental Setup} \label{Implementation}
In our experimental setup, the RSA core runs on a 28nm Zynq-7000 SoC on a Zedboard, which hosts a dual-core ARM Cortex-A9 and programmable logic with 53,200 LUTs. 
 The reference clock varies from 10-100 MHz for different experiments. The ARM processor is programmed for a control application to receive the output of the power monitors, translate it to ASCII and transfer it to the host PC through a UART I/O port for further analysis. The tool-flow to evaluate the SHIELD is shown in Fig.~\ref{fig:Toolflow}. In this work, we estimate the power consumption using the Xilinx Power Estimator not to evade the power constraints proposed by Xilinx. The power traces sampled in power monitors are analyzed using MATLAB scripts for overhead and accuracy analysis.

%% file: Sections/Result.tex
  \subsection{Defense Evaluation} 
To evaluate our defense mechanism, we analyze the distribution of the sampled power trace and compute the attack effort (number of traces) to extract the 1024 bit encryption key of RSA. We perform the SCA by detecting the peaks and valleys in the sampled traces obtained from power monitors. The principles of noise-addition technique, is to increase the power consumption of the chip to obfuscate the behaviour of the victim when it is not using power. In this case, attacker can reveal the processing key bits by observing the values of the sampled power at runtime, because the share of power for attacker circuit changes based on the victim's power consumption.

\begin{figure}[ht]
    \centering
    \includegraphics[width=\linewidth]{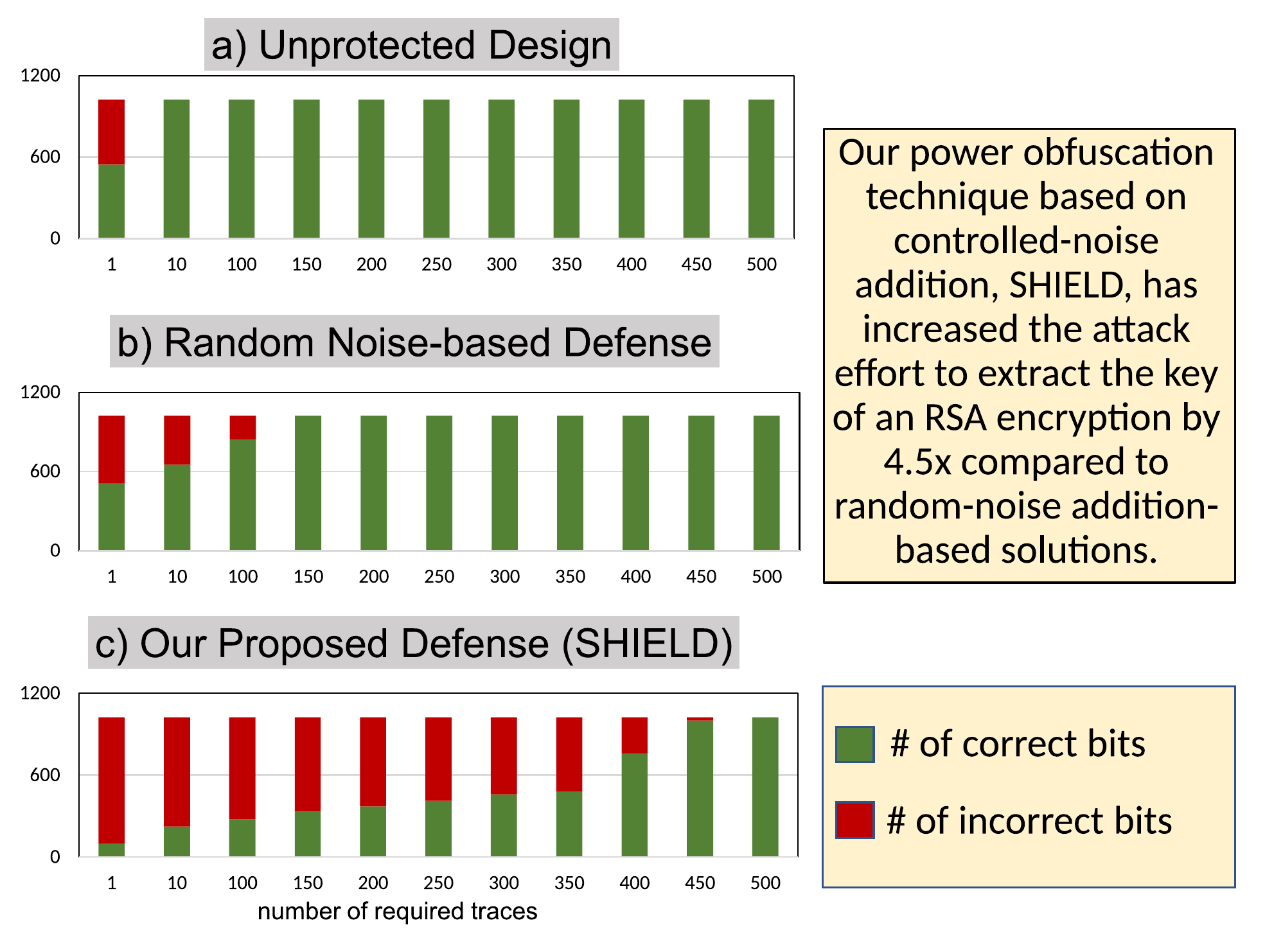}
    \caption{The attack effort to extract the RSA encryption key.}
    \label{fig:accuracy}
\end{figure}

\textbf{Attack Effort (number of traces to extract the key):} 
To analyze the impact of SHIELD on the effort for Simple Power Analysis attack, we compute the number of correct bits and incorrect bits in the extracted key over multiple iterations, as shown in Fig.~\ref{fig:accuracy}. In our attack model, we assume the worst-case situation for the victim, which is that there are no other tenants in the system save the victim and attacker. The inclusion of additional tenants increases the bit error in the observed pattern by the attacker. As a result, we assumed the least effort possible on the part of the attacker in terms of the number of traces required to decrypt 1024-bits of RSA encryption. It is assumed that in a multi-tenant environment, the needed number of traces will increase as a result of noise brought into the system by other tenants' switching activities. We observed a 4.5x increase in attack effort in the presence of SHIELD compared to a random noise addition-based protection and a more than 166x increase in attack effort compared to an unprotected design in this experiment. Thus, SHIELD effectively hides the power's behavior. 
\begin{table}[h]
\centering
\begin{tabular}{l|l|l|l}
Sampling frequency (MHz) & 10 & 50    & 100  \\ \hline
Average of sample counts & 2.00  & 2.88 & 3.28
\end{tabular}
\caption{\centering Average reaction time in terms of number of samples}
\label{tab:reaction}
\end{table}

\textbf{Effectiveness Analysis:} 
To evaluate the SHIELD behavior for different iterations, we compute the a normalized cross correlation of consecutive traces~\cite{corr} in power traces collected from the unprotected design, random noise addition-based, and SHIELD in Fig.~\ref{fig:Corr}. By balancing the power usage, the amplitude of the differential trace is reduced. This is how SHIELD is built to reduce the correlation between distinct power traces; as demonstrated in Section VII, SHIELD cannot avoid DPA but can significantly increase its cost by reducing the correlation between different power traces. We observe that the range of values in first two, are very large as compared to SHIELD. This analysis shows that in the presence of SHIELD, similarity metric for power traces using SHIELD is higher than unprotected and protected with random noise, and this practically hinders the information leakage of different inputs for the attacker. 


  \begin{figure*}[bht]
      \centering
      \includegraphics[width=\textwidth]{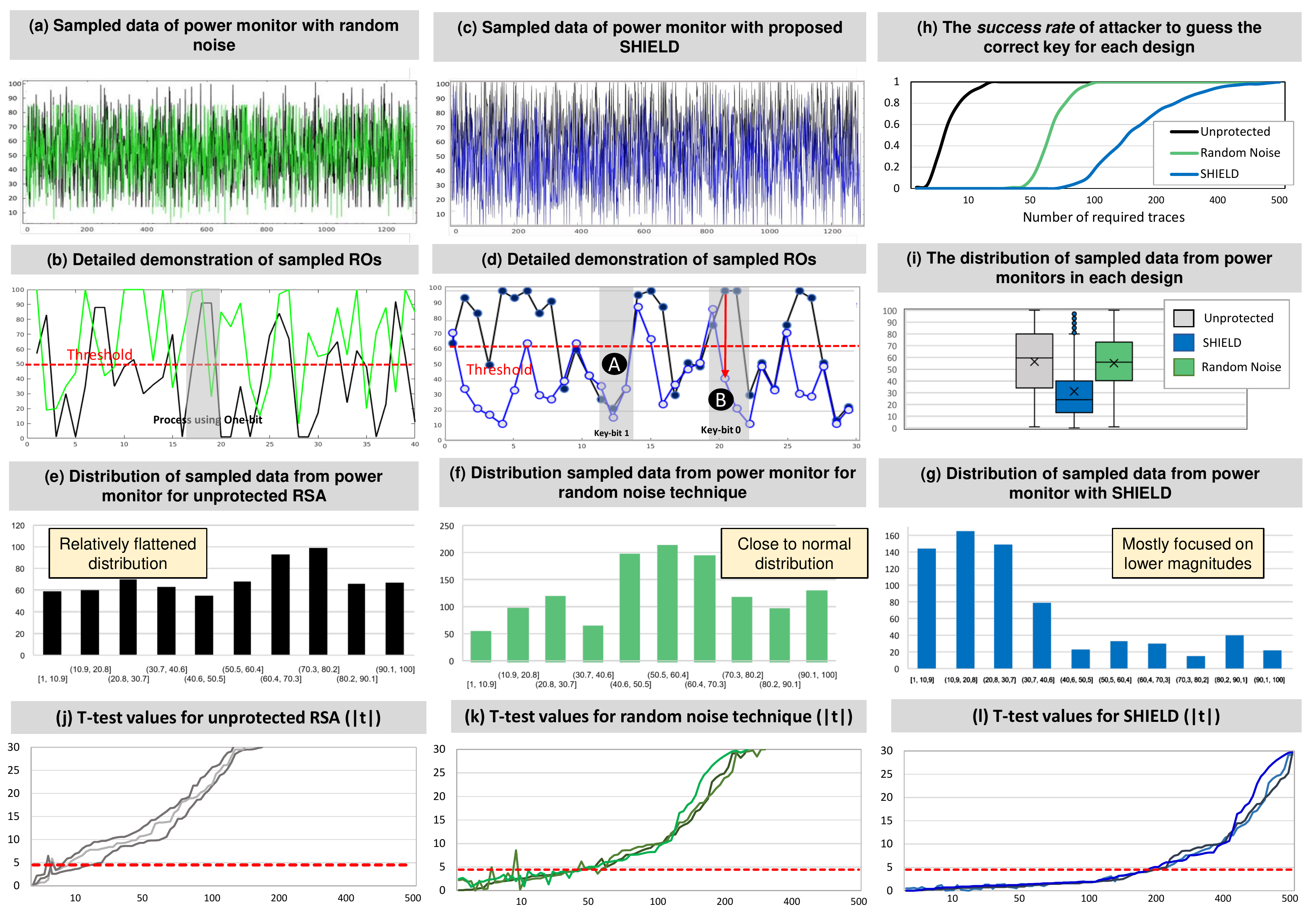}
      \caption{Performance analysis for SHIELD methodology in comparison with original design and random-noise addition.}
      \label{fig:Shift}
  \end{figure*}
\textbf{Analysis of the Power Samples distribution:} Fig.~\ref{fig:Shift} shows the distribution of power consumption over time. In this figure, the sampled data of the power monitors for an unprotected RSA, RSA with random noise addition-based defense, and RSA with SHIELD are shown in black, green, and blue traces, respectively. From this analysis, we made the following key observations:
\begin{enumerate}
    \item The random noise addition-based defense obfuscated a limited number of sampled data (see Fig.~\ref{fig:Shift}(b)) by adding random amount of positive noise to the observed power. By aligning the traces we see while some of the sample points are shifted to lower values (due to less share of power and lower frequency in RO cycle counts) most of the samples are still showing above threshold values. In Fig.~\ref{fig:Shift}(d) it has been shown that our proposed SHIELD has detected most of the peaks in sampled data of power monitors (see~\Circled{A} and~\Circled{B}) and activated appropriate number of ROs to compensate the victim's silent behaviour. The reaction time of the Noise Generators is a function of how many samples it takes to activate the corresponding ROs. The sampling period of RO is controlled by reference frequency and a predefined counter value. In Table.~\ref{tab:reaction} shows the average number of samples as reaction time for different reference frequencies when the frequency of the RSA accelerator is set at 10 MHz.
    \item  The distribution analysis of power monitor values for each iteration of encryption, presents that power trace is more compact in lower values (see Fig.~\ref{fig:Shift}(g)) compared to the distribution of values for unprotected design (see Fig.~\ref{fig:Shift}(e)) and random noise-addition (see Fig.~\ref{fig:Shift}(f)). While random-noise has moved the distribution of values of attack monitors to lower numbers, SHIELD acts more effective using controlled-noise addition (see Fig.~\ref{fig:Shift}(i)). This is due to the lower share of power to ROs and consequently lower frequency of oscillation to be counted.
    \item The success rate in key-stealing attacks is coupled with the attacker's effort to extract the key of encryption. Following the concept of success rate in the literature, we define the n-th order success rate of a side-channel attack as the probability that the target secret is ranked among the n first key guesses by the score vector in Fig.~\ref{fig:Shift}(h)
    \item Test Vector Leakage Assessment: The TVLA methodology searches the leakage points by exploiting Welch t-test~\cite{Welch}, which computes the t-values of runtime power traces. If the TVLA value is within $\pm$4.5, the traces are considered not to contain data-dependent leakage in the intermediate values~\cite{TVLA}. As shown in Fig.~\ref{fig:Shift}(j), the unprotected RSA passes the aforementioned TVLA threshold in $\leq$5 traces, , while in the random noise protected design it happens in $\leq$50 traces Fig.~\ref{fig:Shift}(k), and in SHIELD protected design TVLA threshold passes $\sim$200, shown in Fig.~\ref{fig:Shift}(l). Therefore, employing SHIELD increases the RSA protection against simple power SCAs by $\sim$4$\times$. 
    
\end{enumerate}

\subsection{Implementation Overhead}
We analyzed the area (in terms of number of FFs) and power overheads of SHIELD for different sizes of power monitors (see Fig.~\ref{fig:OA}). From this analysis, we made the following key observations:
\begin{enumerate}
    \item The protected design by SHIELD consumes, on average, 54\% less dynamic power compared to the random-noise protection technique, because of our proposed adaptive activation algorithm for noise generators, (see~\Circled{A} and~\Circled{B} in Fig.~\ref{fig:OA}(a)). It is worth noting that SHIELD has at most ~58\% increase of power consumption in compare to the unprotected design. 
    \item It is demonstrated that, in the best case, 26\% fewer FFs are used as compared to random-noise addition (See~\Circled{E} and~\Circled{F} in Fig.~\ref{fig:OA}(b)). The reason behind this is a lower number of FFs in ROs. 
    \item Accurate monitor and high-frequency oscillation in the noise generator components in SHIELD, results in a faster response to voltage-drop detection in comparison to random noise. Hence, SHIELD increased attack effort to exploit RSA by up to 4.5x compared to random-noise addition (see~\Circled{H} and~\Circled{I} in Fig~\ref{fig:OA}(c)) and 166x increase in number of required traces compared to the unprotected design.
\end{enumerate}

%% file: Sections/Conclusion.tex
 \begin{figure}[ht]
    \centering
    \includegraphics[width= 1.0\linewidth]{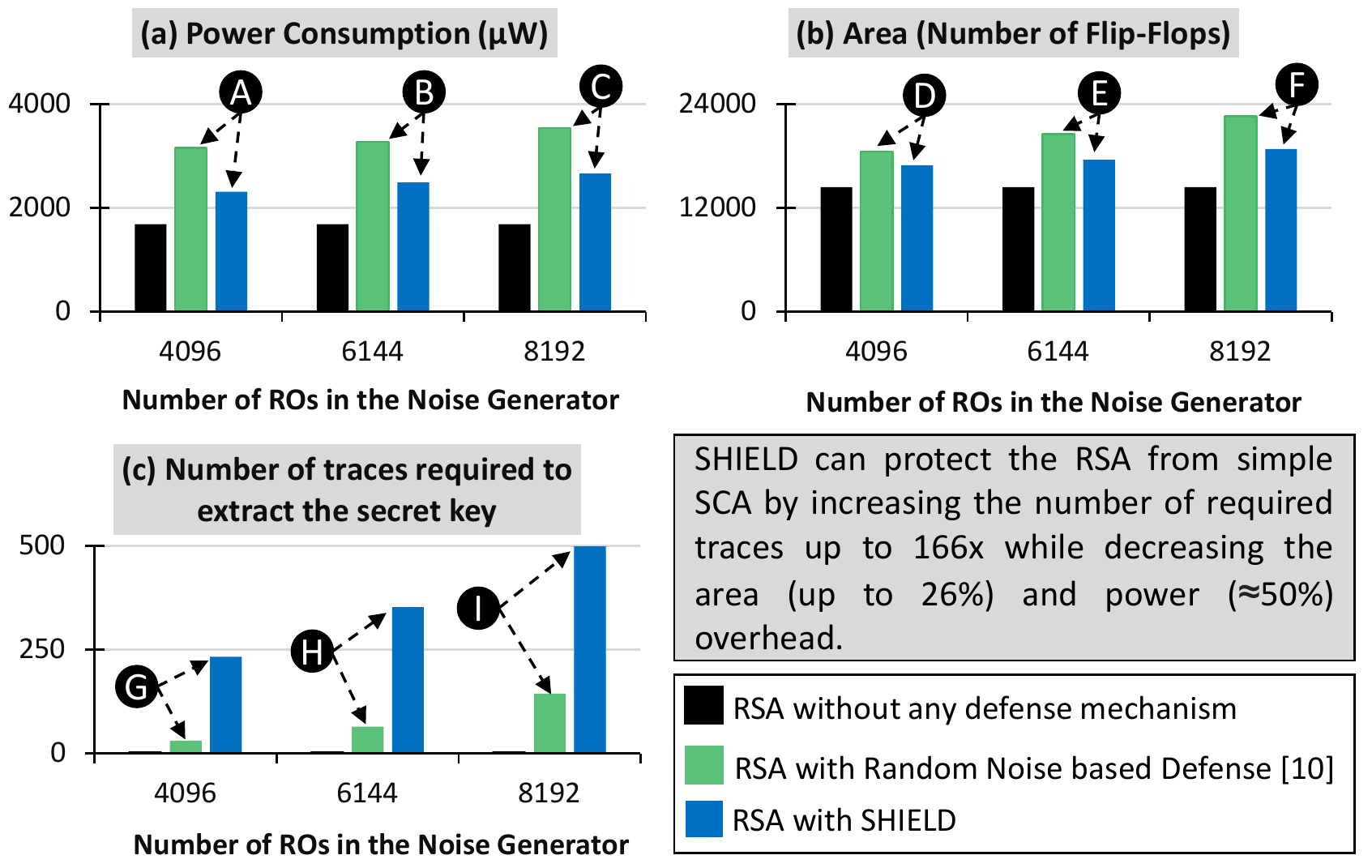}
    \caption{On-chip area and power overhead of our SHIELD and state-of-the-art defense~\cite{zhao_2018} when protecting the RSA design. Note, in this analysis power consumption overhead is calculated using Vivado Power Estimator and the area overhead is reported by Vivado Place\&Route tool.}
    \label{fig:OA} 
    \vspace{-5pt}
\end{figure}
\section{Conclusion}
With the increase in the adoption of cloud-based FPGAs for remote computing, e.g., virtual accelerators, and sharing the infrastructure between different parties, there is a need to research on the security of tenants. Recently, some reports have been published for the side-channel attacks on multi-tenant FPGAs. In this work, we aim to provide a defense mechanism against such attacks. While state-of-the-art proposes obfuscation technique using random-noise addition to power traces, we propose a controlled-noise obfuscation technique, called SHIELD. The proposed active SHIELD benefits from an offline stage to explore design-space for an appropriate application-specific configuration. This power monitor is used to adjust the noise-addition level at run-time to obfuscate the data-dependent power fluctuations in the target application. Our methodology shows up to 54\% less power consumption and up to 26\% less area overhead in comparison to the state-of-the-art techniques. 
